\documentclass{iaus}
\usepackage{graphicx}

\title[Oscillations of supergiants] 
{Radial and nonradial oscillations of massive supergiants}

\author[H. Saio]   
{Hideyuki Saio}

\affiliation{Astronomical Institute, Graduate School of Science, Tohoku University, Sendai, Miyagi 980-8578, Japan
\\ email: {\tt saio@astr.tohoku.ac.jp}}

\pubyear{2010}
\volume{272}  
\pagerange{--}
\jname{Active OB stars: structure, evolution, mass loss and critical limits}
\editors{C. Neiner, G. Wade, G. Meynet \& G. Peter }
\begin{document}

\maketitle

\begin{abstract}
Stability of radial and nonradial oscillations of massive supergiants is 
discussed. The kappa-mechanism and strange-mode instability excite
oscillations having various periods in wide ranges of the upper part 
of the HR diagram. 
In addition, in very luminous ($\log L/L_\odot \gtrsim 5.9$)
models, monotonously unstable modes exist, 
which probably indicates the occurrence of optically thick winds. 
The instability boundary is not far from the Humphreys-Davidson limit. 
Furthermore, it is found that there exist low-degree
($\ell = 1, 2$) oscillatory convection modes associated with the Fe-opacity
peak convection zone, and they can emerge to the stellar surface so that  
they are very likely observable in a considerable range in the HR diagram.
The convection modes have periods similar to g-modes, and their 
growth-times are comparable to the periods.
Theoretical predictions are compared with some of the supergiant variables.
\end{abstract}

\firstsection 

\section{Introduction}
Light and velocity variations on various time-scales are common 
in very luminous stars (e.g., \cite[van Genderen 1989]{vanG89},
\cite[van Leeuwen et al 1998]{vanL98}).
Those variations are caused by various kinds of instabilities.
Here we discuss mainly the cause of microvariations of massive supergiants
based on linear stability analyses applied to evolutionary models 
of massive stars.
The evolution models were calculated by a Henyey-type code 
using OPAL opacity tables
(\cite[Iglesias \& Rogers 1996]{opal}). 
Wind mass-loss is included for the evolutionary models for
$M_i\gtrsim 30M_\odot$ 
($M_i$=initial mass) based on the mass-loss rates of \cite{vink01}.
Linear stability analyses were performed using the methods given
in \cite{sa83} for radial modes and \cite{sa80} for nonradial modes,
where the outer mechanical boundary condition was modified to
$\delta P_{\rm gas} \rightarrow 0$ ($\delta P_{\rm gas}$ means the 
Lagrangean perturbation of gas pressure) taking into account the fact
that radiation pressure is dominant near the outer boundary.
It should be noted, however, that the effect of winds on radial and
nonradial oscillations is not included because the effect is not
well understood yet. 

\section{Stability of radial modes}

\begin{figure}[t]
\begin{center}
 \includegraphics[width=2.9in]{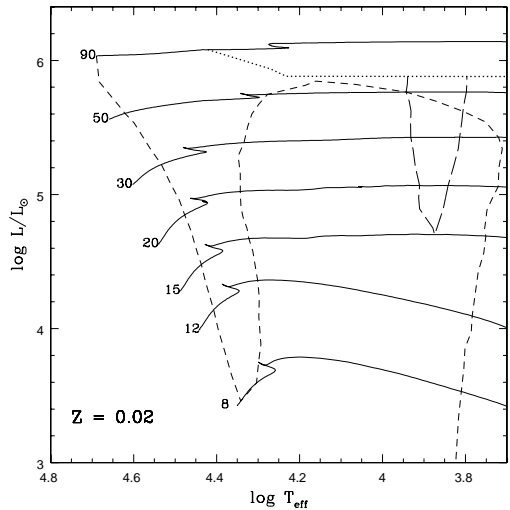} 
 \caption{Instability boundaries of radial modes 
and selected evolutionary
tracks. Short- and long-dashed lines indicate stability boundaries
for low-order and relatively high-order radial modes, respectively.
Dotted line indicates the boundary for monotonously unstable radial modes.
Numbers along evolutionary tracks indicate initial masses in solar
units.
The instability boundaries for radial modes also represent approximately
the boundaries for low-degree ($\ell \lesssim 2$) nonradial p-modes.}
   \label{fig:rad}
\end{center}
\end{figure}

Figure~\ref{fig:rad} shows instability boundaries of spherical symmetric modes
and selected evolutionary tracks.
Short-dashed line indicates the instability boundary for low-order modes.
The nearly vertical ``finger'' ($4.4 \gtrsim\log T_{\rm eff} \gtrsim 4.3$) is 
the well-known $\beta$ Cep instability strip,
in which low-order radial modes and nonradial p-modes are excited by the 
kappa-mechanism  at the Fe-peak of opacity around $T\sim2\times10^5$K.
At the luminous part of the instability strip, the cool-side boundary 
bents to become horizontal around $\log(L/L_\odot) \sim 5.8$. 
This is due to the strange-mode instability which occurs in models with
sufficiently high luminosity to mass ratios as 
$L/M \gtrsim 10^4 L_\odot/M_\odot$.
The properties of the strange modes have been investigated by e.g.,
\cite{gl93}, \cite{gl94}, \cite{sa98} (see also \cite[Saio 2008]{sa08}).

Nearly vertical part of the boundary around $\log T_{\rm eff} \sim 3.8$ 
indicates the well-known blue
edge of the Cepheid instability strip. (Red edge is not obtained because
the perturbation of convective flux is neglected in this analysis.)

In the vertical narrow region indicated by long-dashed line around
$\log T_{\rm eff}\sim 3.9 - 3.8$, relatively high-order radial modes
are excited around hydrogen ionization zone. (Low-degree
nonradial modes with similar frequencies are also excited.)
The amplitude of these modes are extremely confined to the outermost
layers above the hydrogen ionization zone.
Since these modes exist even under the NAR approximation where
thermal-time is set to be zero (\cite[Gautschy \& Glatzel 1990]{gg90}),
they may be classified as strange modes.
These modes have got little attention so far 
(cf. \cite[Gautschy 2009]{afg09}). 

Dotted line in Fig.~\ref{fig:rad} shows the boundary above which
monotonously unstable modes exist. The growth times of these modes
are much shorter than the timescale of evolution.
The presence of such monotonously unstable modes have not been 
recognized before.
Such a mode probably corresponds to the presence
of an optically thick wind as investigated
by \cite{ka92} for WR stars.
It is interesting to note that the boundary is not far from the
Humphreys-Davidson limit (\cite[Humphreys \& Davidson 1979]{hd79}).

\section{Stability of nonradial modes}

\begin{figure}[t]
\begin{center}
 \includegraphics[width=2.9in]{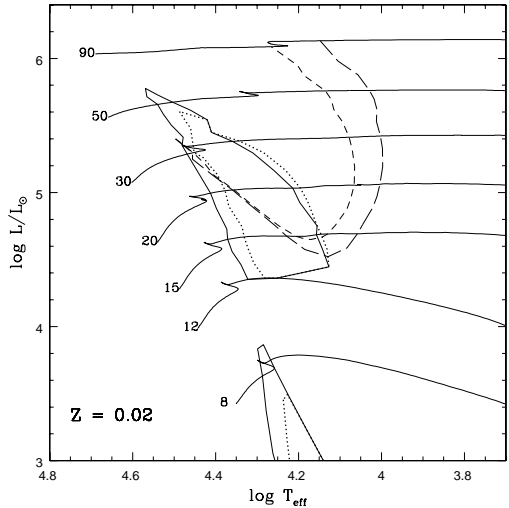} 
 \caption{Ranges where nonradial oscillations are expected to be detected.
The instability boundaries for SPB-type low-degree high-order g-modes 
are shown by solid ($\ell = 2$) and dotted ($\ell = 1$) lines. 
The ranges where oscillatory convection modes should be
observable are indicated by long- and short-dashed lines for
$\ell = 1$ and $\ell=2$, respectively.
}
   \label{fig:nonrad}
\end{center}
\end{figure}

The instability ranges of low-degree high-order g-modes 
are shown in Fig.~\ref{fig:nonrad} by solid and dotted lines for
$\ell=2$ and $\ell=1$ modes, respectively.
These modes are excited by the kappa-mechanism at the Fe opacity
peak. They are responsible for the long-period variations in 
slowly pulsating B (SPB) stars.
Such g-modes can be excited even in supergiants (SPBsg)
because a shell convection zone associated with hydrogen burning
reflects some g-modes and hence suppresses dissipation otherwise
expected in the core (\cite[Saio et al 2006]{sa06}). 

The long dashed and short dashed lines in Fig.~\ref{fig:nonrad} indicate
ranges where oscillatory convection modes of $\ell=1$ and $\ell=2$,
respectively, are expected to be observable.
It is well known that linear convection modes are monotonously (dynamically) 
unstable in {\it adiabatic analyses}. \cite{sh81} found, however, that   
the high-degree ($\ell \ge 10$) convection modes become overstable 
(oscillatory) when the nonadiabatic effect is included in 
luminous ($L/L_\odot =10^5$) models hotter than the cepheid instability
strip.

In our massive star models, 
it is found that low-degree ($\ell\le2$) oscillatory convection modes
exist associated with the Fe-opacity peak convection zones,
and some of these modes are expected to be observable.
Since the growth time of a convection mode is short (comparable
to the period), the mode is expected to
have a large amplitude in the convective zone.
Therefore, the visibility of the oscillatory convective modes  
can be measured by the ratio of the photospheric amplitude 
to the maximum amplitude
in the interior (mostly in the convection zone).
Assuming that an oscillatory convection mode is observable 
when the ratio is larger than 0.2, the boundaries of the visible
ranges are shown in Fig.~\ref{fig:nonrad} by dashed lines.
Oscillatory convection modes are visible in sufficiently luminous 
($\log L/L_\odot \gtrsim 4.6$) B-type stars. 
Although there are many oscillatory convection modes in a star, 
only one or two modes for a given $\ell$ are visible because the
other modes are well confined to the convection zone.
Periods of these modes are comparable to g-modes much longer
than those of radial modes. 
These oscillatory convection modes might be responsible for long-period
variations in supergiant stars.

\section{Comparison with supergiant variables}

\begin{figure}[t]
\begin{center}
 \includegraphics[width=3.4in]{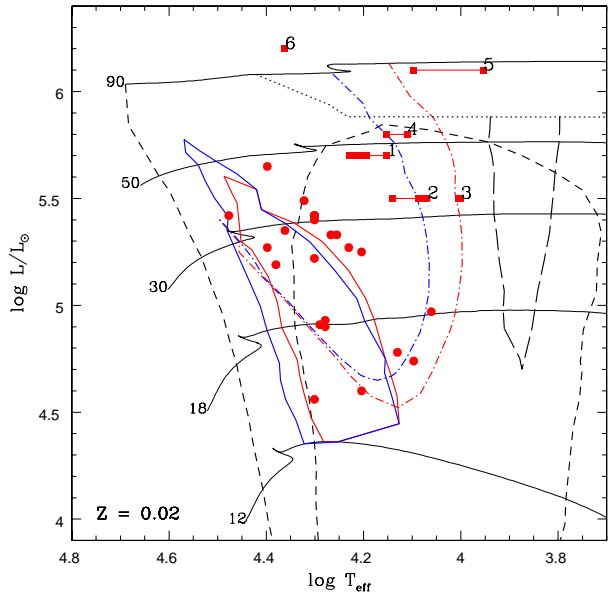} 
 \caption{
Ranges where (quasi-)periodic variable stars are expected
on the HR diagram
are compared with supergiant variables analyzed by \cite{Lef07} 
(big filled circles) and LBVs by \cite{Lam98} (filled squares).
Positions at different LBV (or S Dor) phases for each star 
are connected with thin lines; 1 = R 712, 2 = HR Car, 3 = 164 G Sco,
4 = S Dor, 5 = R 1273, 6 = AG Car. 
}
   \label{fig:hrd}
\end{center}
\end{figure}

\begin{figure}[th]
\begin{center}
 \includegraphics[width=3.3in]{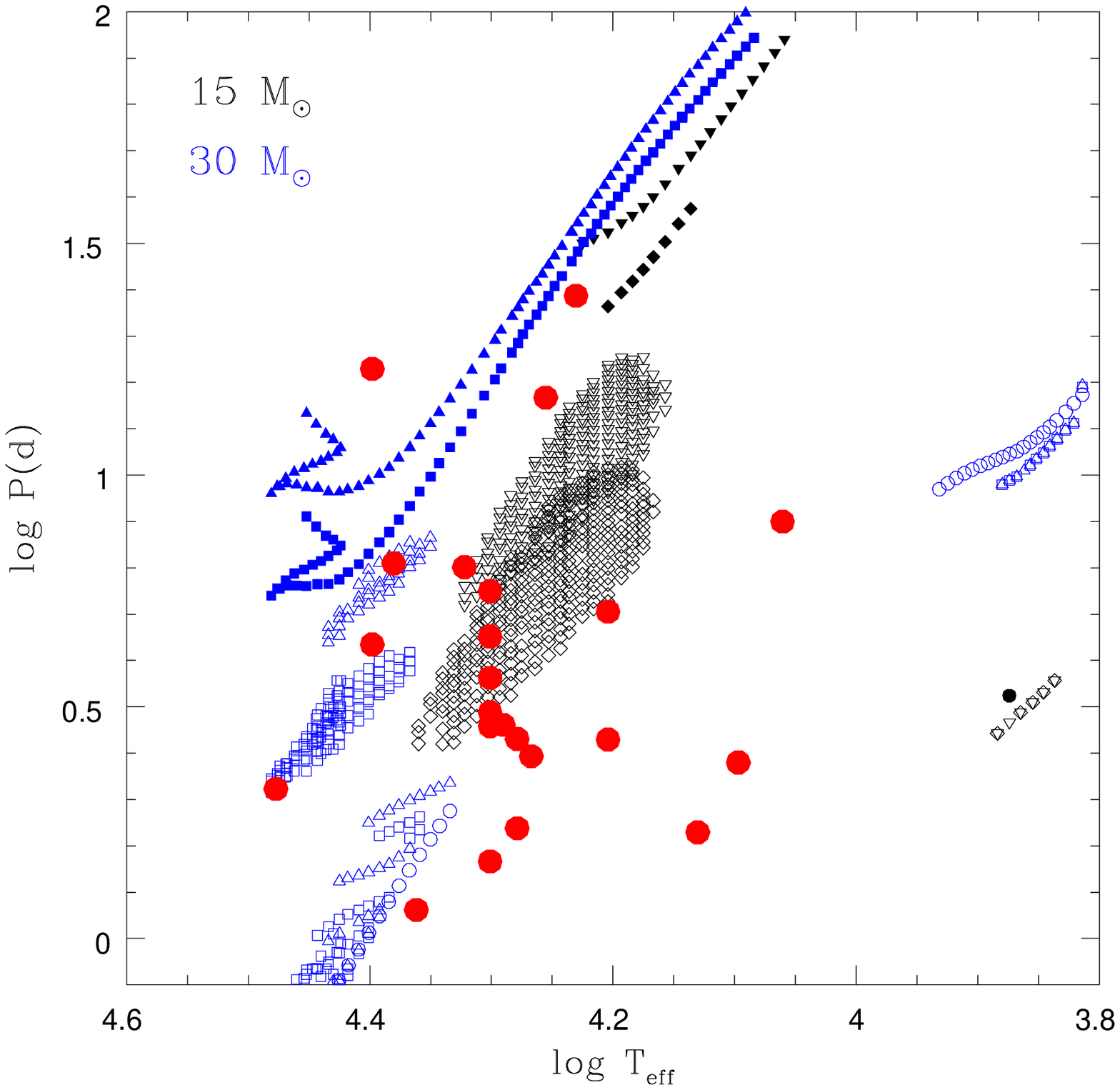} 
 \caption{
Periods of excited modes and of observable oscillatory convection
modes versus effective temperature, where open ($30 M_\odot$) and 
filled ($15 M_\odot$) circles are radial modes;
triangles ($30 M_\odot$) and inverted triangles ($15 M_\odot$) 
are $\ell=1$ modes; squares ($30 M_\odot$) and diamonds ($15 M_\odot$) 
are $\ell=2$ modes.
Convection modes are shown by filled symbols for nonradial modes.
Big dots show observed periods-$T_{\rm eff}$ relations of supergiants 
from \cite{Lef07}.
}
   \label{fig:teperi_low}
\end{center}
\end{figure}

\begin{figure}[th]
\begin{center}
 \includegraphics[width=3.3in]{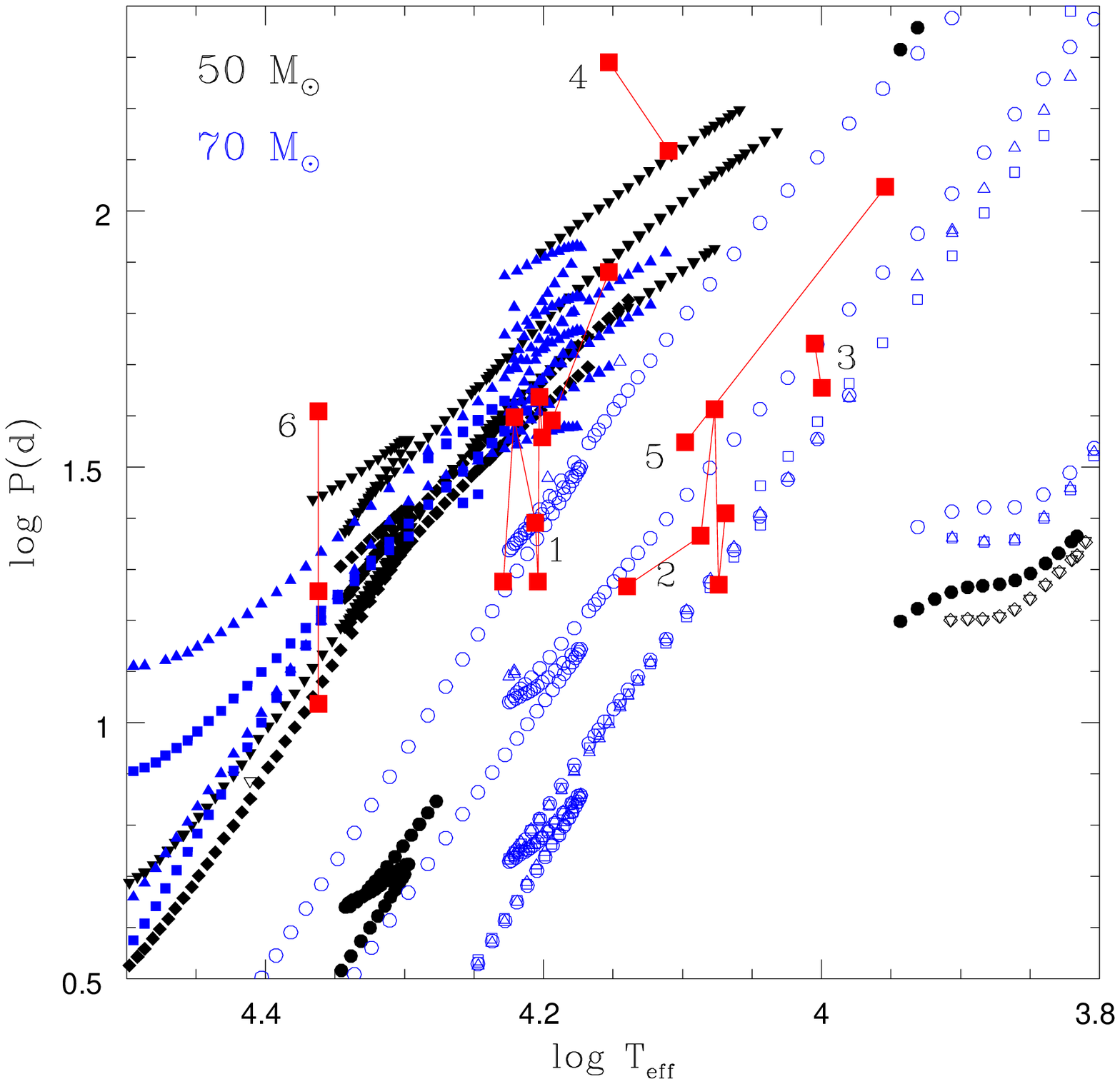} 
 \caption{
Theoretical periods of excited modes and of observable oscillatory convection
modes versus effective temperature, where open ($70 M_\odot$) and filled 
($50 M_\odot$) circles are radial modes;
triangles ($70 M_\odot$) and inverted triangles ($50 M_\odot$) 
are $\ell=1$ modes;
and squares ($70 M_\odot$) and diamonds ($50 M_\odot$) are $\ell=2$ modes.
Convection modes are shown by filled symbols for nonradial modes.
Big filled squares connected with thin lines show observed
periods-$T_{\rm eff}$ relations of LBVs obtained by \cite{Lam98}.
Periods at different LBV (or S Dor) phases for each star 
are connected with thin lines; 1 = R~712, 2 = HR~Car, 3 = 164~G~Sco,
4 = S~Dor, 5 = R~1273, 6 = AG~Car. 
}
   \label{fig:teperi}
\end{center}
\end{figure}

Observed positions of variable supergiants on the HR diagram are
compared with the instability and visibility boundaries in Fig.~\ref{fig:hrd}.
Their periods-$T_{\rm eff}$ relations are
compared with theoretical ones in Figs.~\ref{fig:teperi_low} 
and~\ref{fig:teperi}.
Fig.~\ref{fig:hrd} indicates that all the hotter 
($\log T_{\rm eff}\gtrsim 4$) and luminous ($\log L/L_\odot \gtrsim 4.5$)
stars are expected to show (quasi-)periodic variations,
which is consistent with the observational fact  that
no or at most a very little number of stable supergiants exist in a
spectral range of O9 -- A0 as found by \cite{vanG89}.

Big dots in Fig.~\ref{fig:hrd} are relatively less luminous supergiant 
variables analyzed by \cite{Lef07}. 
This figure indicates most of them to have
masses ranging from $\sim 15M_\odot$ to $\sim 30M_\odot$. 
They are located on the HR diagram in the g-mode instability regions
or visible range of oscillatory convection modes.
Fig.~\ref{fig:teperi_low} compares the periods of these stars
as function of the effective temperature with theoretical ones of
$15M_\odot$ and $30M_\odot$ models.
This figure indicates that for most of these stars periods seem consistent to
low-degree high-order g-modes (SPBsg) or oscillatory convection modes.
We note, however, that for the coolest three stars the periods 
are shorter than any of the excited modes.
Although these three stars are located on the HR diagram in the region where  
oscillatory convection modes should be visible
(Fig.~\ref{fig:hrd}), the periods are much shorter than those of
the oscillatory convection modes.
If these effective temperatures are accurate, an unknown excitation mechanism
might be working in these stars. 

Fig.~\ref{fig:teperi} compares theoretical periods of very massive
models ($M_i=50M_\odot$ and $70M_\odot$) with periods of microvariations
of some LVB stars analyzed by \cite{Lam98}, each of which has different
periods and effective temperature depending on the LBV (S Dor) phases.
Figs.~\ref{fig:teperi} and~\ref{fig:hrd} indicate that
the microvariations of  R~712 (1), S~Dor (4) and AG~Car (6) 
are consistent to the properties of oscillatory convection modes.

The periods of microvariations of HR~Car (2), 164~G~Sco (3), and AG~Car (5) 
are consistent with periods of strange modes.
However, luminosities of HR~Car and 164~G~Sco are too low for 
the strange modes to exist (Fig.~\ref{fig:hrd}). 
Further investigations are needed to resolve the discrepancy.

\section{Summary}
We have discussed various instabilities that occur in massive supergiants.
Radial modes and nonradial p- and g-modes are excited by the 
kappa-mechanism at the Fe opacity bump at $T\sim2\times10^5$.

In a star with a very high luminosity to mass ratio of 
$L/M \gtrsim 10^4L_\odot/M_\odot$, the strange mode
instability works for radial and nonradial modes.
Strange modes seem to be
responsible for quasi-periodic variations in some of the luminous supergiants. 

In addition, it is found that in very luminous models 
($\log L/L_\odot \gtrsim 5.9$) a monotonously unstable radial mode 
exists, which is probably related to the occurrence of an optically thick
wind. It is interesting to note that the instability boundary 
roughly coincides with the Humphreys-Davidson limit.

Furthermore, we found that low-degree ($\ell=1, 2$) oscillatory convection 
modes exist in the convection zones caused by the Fe opacity peak, and
that some of them can emerge to the stellar surface and hence be observable.
The oscillatory convection modes have periods of $10 \sim 10^2$days
depending of the effective temperature, which are longer than those of
strange modes.  The growth-times are comparable to the periods.
They seem to be consistent with the properties of long-period microvariations
in LVB stars (see \cite[Saio 2010]{sa10} for further discussions).  
  

\end{document}